\documentclass[twocolumn,groupedaddress,showpacs,nofootinbib]{revtex4}
\usepackage{graphicx}
\usepackage{amsmath}
\usepackage{epsfig}
\newcommand {\ignore}[1]{}
\renewcommand{\baselinestretch}{1.15}
\def\e6{$\mathrm{E(6)}$ }
\def\10{$\mathrm{SO(10)}$ }
\def\21{$\mathrm{SU(2) \otimes U(1)}$ }
\def\31{$\mathrm{SU(3) \otimes U(1)_Q}$ }

\def\3211{$\mathrm{SU(3) \otimes SU(2)_L \otimes U(1)_R \otimes U(1)_{B-L}}$ }
\def\422{$\mathrm{SU(4) \otimes SU(2)_L \otimes SU(2)_R}$ }

\let\vev\VEV
\def\lsim{\raise0.3ex\hbox{$\;<$\kern-0.75em\raise-1.1ex\hbox{$\sim\;$}}}
\def\gsim{\raise0.3ex\hbox{$\;>$\kern-0.75em\raise-1.1ex\hbox{$\sim\;$}}}

\newcommand{\AddrAHEP}{%
  AHEP Group, Institut de F\'{\i}sica Corpuscular --
  C.S.I.C./Universitat de Val{\`e}ncia \\
  Edificio Institutos de Paterna, Apt 22085, E--46071 Valencia, Spain}

\begin{document}

\title{Neutrino masses, leptogenesis and dark matter in hybrid seesaw}

\author{Pei-Hong Gu$^{1}_{}$}
\email{pgu@ictp.it}

\author{M. Hirsch$^{2}_{}$}
\email{mahirsch@ific.uv.es}

\author{Utpal Sarkar$^{3}_{}$}
\email{utpal@prl.res.in}\homepage{http://www.prl.res.in/~utpal}

\author{J.W.F. Valle$^{2}_{}$}
\email{valle@ific.uv.es}\homepage{http://ahep.uv.es}

\affiliation{$^{1}_{}$The Abdus Salam International Centre for
Theoretical Physics, Strada Costiera 11, 34014 Trieste, Italy\\
$^{2}_{}$ \AddrAHEP \\
$^{3}_{}$Physical Research Laboratory, Ahmedabad 380009, India}

\begin{abstract}

  We suggest a hybrid seesaw model where relatively ``light''
  right-handed neutrinos give no contribution to the neutrino mass matrix
  due to a special symmetry. This allows their Yukawa couplings to the
  standard model particles to be relatively strong, so that the
  standard model Higgs boson can decay dominantly to a left and a
  right-handed neutrino, leaving another stable right-handed neutrino
  as cold dark matter.  In our model neutrino masses arise via the
  type-II seesaw mechanism, the Higgs triplet scalars being also
  responsible for the generation of the matter-antimatter asymmetry
  via the leptogenesis mechanism.

\end{abstract}

\maketitle

\section{Introduction}

Laboratory experiments with reactors and accelerator neutrinos have
confirmed the observations of solar and atmospheric
neutrinos~\cite{Neutrino2008}, establishing the phenomenon of neutrino
oscillations and hence the need for small but nonzero neutrino
masses~\cite{Schwetz:2008er}. Since neutrinos are massless in the
$SU(3)_{c}^{}\times SU(2)_{L}^{}\times U(1)_{Y}^{}$ Standard Model
(SM) this implies the need for new physics, whose detailed nature
constitutes one of our deepest current challenges in particle
physics~\cite{Valle:2006vb}. The simplest extension of the SM to
explain the neutrino masses is to include singlet right-handed
neutrinos and/or triplet scalars~\cite{schechter:1980gr}. The
inclusion of the former is justified by the fact that there are
right-handed partners for all other fermions of the SM. The singlet
right-handed neutrinos will in general have Majorana masses as well as
a Dirac mass term, of the order of the charged fermion masses. The
latter arise from the usual Yukawa interaction between left- and
right-handed leptons, once the SM Higgs doublet acquires a vacuum
expectation value ($vev$).
The terms involving the right-handed neutrinos are,
\begin{eqnarray}
\label{eq:lagrangian1}
    {\cal L} \supset -\frac{1}{2}(M_{N}^{})_{ij}^{} \overline{N_{R_i}^{c}}
    N_{R_j}^{}
    -h_{ij } \overline{\ell_{L_i}^{}} \phi N_{R_j}^{} + \textrm{H.c.} \,.
\end{eqnarray}
Here $N_{R}$, $\ell_{L}$ and $\phi$ denote the right-handed neutrinos,
the left-handed leptons and Higgs doublet, respectively. In the basis
$(\nu_{L}^{},~ N_R^c) $, the neutrino mass matrix reads,
\begin{equation}
\label{eq:matrix}
    M_\nu^{} = \begin{pmatrix} 0 &  m_D^{} \cr m_D^{T} & M_N^{}
\end{pmatrix} \,,
\end{equation}
where $(m_{D}^{})_{ij}^{} = h_{ij }^{}\vev{\phi}$ with
$\vev{\phi}\simeq 174\,\textrm{GeV}$. After block-diagonalization, one
gets
\begin{equation}
    m_\nu^{} \simeq -m_D^{} \frac{1}{M_N^{}} m_D^T \,.
\end{equation}
If the Majorana masses of the right-handed neutrinos are much larger
than the Dirac masses, the left-handed neutrinos will naturally
acquire a tiny Majorana through the type-I seesaw
mechanism~\cite{schechter:1980gr}~\cite{Minkowski:1977sc,schechter:1982cv}.
In order to account for neutrino masses in the eV range one needs,
\begin{equation}
h_{ij}^{} h_{kj}^{} (M_N^{})_{jk}^{-1}  \sim 10^{-23}~{\rm GeV}^{-1}
\,. \label{eq:con1}
\end{equation}
so that the Majorana masses $M_N^{}$ are required to be orders of
magnitude larger than the electroweak symmetry breaking scale
$\sim\vev{\phi}$, unless the coefficients $h_{ij }^{}$ are very small,
\begin{equation}
h_{ij} \lsim 10^{-11} \,. \label{eq:con2}
\end{equation}

There is an attractive way to avoid this conclusion in the framework
of the inverse seesaw
model~\cite{mohapatra:1986bd,gonzalez-garcia:1989rw,Deppisch:2005zm}.
In this case the right-handed neutrinos pair-off with extra singlet
leptons to form Dirac-type neutral heavy leptons in such a way that
their mixing with the doublet neutrinos does not lead to light
neutrino masses in the limit of conserved lepton number. As a result
of this ``symmetry-protection'' right-neutrinos can lie at the TeV
scale and produce signals at
accelerators~\cite{Dittmar:1990yg,delAguila:2008hw}, without conflict
with the observed smallness of neutrino masses.

Here we focus on a seesaw mechanism without additional fermion degrees
of freedom. If one assumes ``generic'' violation of lepton number
through right-handed neutrino Majorana masses, acceptable neutrino
masses require very tiny effective Yukawa couplings connecting the
right-handed neutrinos with the left-handed neutrinos,
Eqs.~(\ref{eq:con1}) and (\ref{eq:con2}), hence the right-handed
neutrinos can not be produced at the LHC, nor can the Higgs boson
decay into neutrinos.
We explore the possibility that the neutrino mass $m_{\nu}^{}$ can be
exactly zero even if the Majorana mass $M_N^{}$ and the Dirac mass
$m_{D}^{}$ are both
non-vanishing~\cite{Kersten:2007vk,deGouvea:2007uz}, as a result of a
cancellation. We present a model where the type-I seesaw contribution
to the neutrino masses vanishes identically due to a suitable
symmetry, avoiding the main constraints on $M_N^{}$ and
$m_{D}^{}$. Neutrino masses arise from the type-II triplet seesaw
mechanism~\cite{schechter:1980gr}.  As a result, relatively light
right-handed neutrinos could have sizeable Yukawa couplings. If the
Majorana masses of the right-handed neutrinos are of the order of TeV
or less, they may be produced at the LHC. We also note that, if it is
heavier than the right-handed neutrino, the Higgs boson in this
scenario will also have a distinct decay channel into a left plus a
right-handed neutrino. Our explicit symmetry will also protect one of
the right-handed neutrinos from decaying, as it has no SM Yukawa
interactions. This naturally accounts for a stable cold dark matter
candidate~\cite{Kim:2008pp,Gu:2007gy}. Finally, our scenario for the
origin of neutrino masses also provides successful
leptogenesis~\cite{fukugita:1986hr,kuzmin:1985mm} induced by the
out-of-equilibrum decays of the heavy scalar
triplets~\cite{PhysRevLett.80.5716}.

\section{Neutrino Mass Matrices}

We shall first discuss the structure of the Dirac and Majorana masses
of the neutrinos and then present the model in detail.  To demonstrate
the basic idea, consider a two-generation scenario in which the type-I
seesaw matrix takes a particular form,
\begin{equation}
M_\nu =
\begin{pmatrix}
0 & 0 & a & a \cr 0 & 0 & b & b \cr
a & b & M & 0 \cr a & b & 0 & -M
\end{pmatrix}\,,
\end{equation}
with $M \gg a,b$. Indeed, in this example, the two left-handed
neutrinos remain massless despite the coexistence of Dirac and
Majorana mass terms, as a result of the cancelation between the
contributions from the two right-handed neutrinos. The latter may be
due to some underlying symmetry.  In this case the Yukawa couplings
thus can be large even if the right-handed neutrinos have ``low''
Majorana masses.

Consider the mass matrix
\begin{eqnarray}\label{eq:matrix2}
M_\nu=
\begin{pmatrix}
0 & 0 & 0 & a \cr 0 & 0 & 0 & b \cr
0 & 0 & 0 & M \cr a & b & M & 0
\end{pmatrix}
\end{eqnarray}
It is easy to see that (i) this matrix reduces to the previous after
diagonalizing out the right handed states and (ii) it emerges from a
$Z_3$ symmetry
\begin{equation}
N_{R_1}^{} \to \omega^2 N_{R_1}^{}\,,~~~~
N_{R_2}^{} \to \omega N_{R_2}^{}\,, ~~~~
\nu_{L_i}^{} \to \omega \nu_{L_i}^{},
\end{equation}
where $\omega$ is the cube-root of 1, $\omega^3 = 1$ and $1 + \omega +
\omega^2 = 0$. All other charged fermions also transform under this
$Z_3$ symmetry as $f \to \omega f$ and the Higgs doublet $\phi$ is
invariant, so that all the usual couplings allowed in the SM remain
the same.  The Lagrangian contributing to the neutrino masses is then
given by
\begin{equation}
{\cal L} \supset - M \overline{N_{R_1}^{c}} N_{R_2}^{} - h_{i2}^{}
\overline{\ell_{L_i}^{}} \phi N_{R_2}^{}+\textrm{H.c.}\,.
\end{equation}
which clearly leads to the form in Eq.~(\ref{eq:matrix2}) once the
$vev$ $\vev{\phi}$ is generated.
This symmetry makes the light neutrinos remain exactly massless.  In
order to generate the required neutrino masses one may either break
this $Z_3$ softly or, alternatively, introduce triplet scalars for
implementing the type-II seesaw~\cite{schechter:1980gr}, without
affecting the symmetry in the right-handed neutrino sector.

This mass matrix can be generalized to three generations.  Consider
the $ 3 \times 3$ mass matrix
\begin{eqnarray}
{\cal M}_\nu &=& \begin{pmatrix} 0 & 0 & 0 & a & a & 0 \cr 0 & 0 & 0
& b & b & 0 \cr 0 & 0 & 0 & c & c & 0 \cr a & b & c & M & 0 & 0 \cr
a & b & c & 0 & -M & 0 \cr 0 & 0 & 0 & 0 & 0 & \tilde M
\end{pmatrix}\,,
\end{eqnarray}
with $M, \tilde M \gg a,b,c$. Clearly, the three left-handed neutrinos
remain exactly massless, while the masses of the right-handed
neutrinos are $+M, -M, \tilde M$ can be ``low'' enough to be
accessible to the LHC.  Note, however, that the third-generation
right-handed neutrino decouples. In another basis of the right-handed
neutrinos one has,
 \begin{eqnarray}
{\cal M}_\nu &=& \begin{pmatrix} 0 & 0 & 0 & a & 0 & 0 \cr 0 & 0 & 0
& b & 0 & 0 \cr 0 & 0 & 0 & c & 0 & 0 \cr a & b & c & 0 & M & 0 \cr
0 & 0 & 0 & M & 0 & 0 \cr 0 & 0 & 0 & 0 & 0 & \tilde M
\end{pmatrix}\,.
\end{eqnarray}
This mass matrix could emerge from a $Z_3$ symmetry
\begin{eqnarray}
& N_{R_1}^{} \to \omega N_{R_1}^{}\,,~~N_{R_2}^{} \to \omega^2
N_{R_2}^{}\,,~~N_{R_3}^{} \to N_{R_3}^{}\,,&\nonumber\\
&\nu_{L_i}^{} \to \omega \nu_{L_i}^{}\,.&
\end{eqnarray}
or, alternatively, from an $U(1)$ global symmetry.  The Lagrangian
containing the right-handed neutrinos,
\begin{eqnarray}
\label{eq:lagrangian2} {\cal L} \supset -M \overline{N_{R_1}^{c}}
N_{R_2}^{} -\frac{1}{2} \tilde M \overline{N_{R_3}^{c}} N_{R_3}^{} -
h_{i1}^{} \overline{\ell_{L_i}^{}} \phi N_{R_1}^{}+\textrm{H.c.}\,.
\end{eqnarray}
implies, in the limit $M \gg h_{i1}^{}\vev{\phi}$, two degenerate
right-handed neutrinos
\begin{eqnarray}
\label{eq:N12}\frac{1}{\sqrt{2}}(N_{R_1}^{}\pm N_{R_2}^{})\rightarrow
N_{R_{1,2}}^{}
\end{eqnarray}
with masses $\pm M$ (opposite CP signs). For $M \lesssim
1\,\textrm{TeV}$ these two states $N_{R_{1,2}}^{}$ would be accessible
to LHC searches. On the other hand, a more interesting possibility may
open up when the Higgs boson is heavier than $N_{R_{1}}^{}$ and
$N_{R_{2}}^{}$. Given the mild constraint on the Dirac mass term, the
Yukawa couplings could be large enough that the Higgs boson would
dominantly decay into a left-handed neutrino and a right-handed one,
posing a new challenge to the Higgs search program at the LHC. The
third right-handed neutrino $N_{R_3}^{}$ has mass $\tilde{M}$ and no
decay modes. Hence it could serve as the dark matter if its relic
density is consistent with the cosmological observations.

\section{Leptogenesis, Neutrino Mass and Dark Matter}

We now propose a realistic model, which contains the previous
phenomenology of the right-handed neutrinos and explains the
matter-antimatter asymmetry, the neutrino masses and the dark
matter. We extend the SM by including three right-handed neutrinos,
two triplet Higgs scalars $\xi_{1,2}^{}\equiv
(\textbf{1},\textbf{3},2)$ and two singlet scalar fields $\sigma,\chi
\equiv (\textbf{1},\textbf{1},0)$. In addition to the SM gauge
symmetry, we also impose a global $U(1)_{lep}^{}$ symmetry of lepton
number, under which the different fields transform as:
\begin{table}[!ht]
\begin{center}
\begin{tabular}{||c|c|c|c|c|c|c|c|c||}
\hline \hline &&&&&&&&\\[-.05in]
$N_{R_1}^{}$ & $N_{R_2}^{}$ & $N_{R_3}^{}$ & $\ell_{L_i}^{}$&
$\xi_{1}^{}$ & $\xi_{2}^{}$&
$\sigma$ & $\chi$ & $\phi$ \\[.05in]
\hline &&&&&&&&\\[-.05in]
1& 3& 2& 1&--2&--1&--4&1&0 \\[.05in]
\hline \hline
\end{tabular}
\end{center}
\caption{ Lepton number assignments. For simplicity, we do not show the
  right-handed charged leptons, which carry the same lepton number as
  their left-handed partners.}
\end{table}

The relevant part of the Lagrangian is given as,
\begin{eqnarray}
\label{eq:lagrangian3} {\cal L} &\supset& -\alpha_1^{} \sigma
\overline{N_{R_1}^{c}} N_{R_2}^{} - \frac{1}{2}\alpha_2^{} \sigma
\overline{N_{R_3}^{c}} N_{R_3}^{}-h^{}_{i1} \overline{\ell_{L_i}^{}}
\phi N_{R_1}^{}
\nonumber \\
&&-\frac{1}{2} f_{ij} \overline{\ell_{L_i}^{c}} i\tau_2^{}\xi_{1}^{}
\ell_{L_j}^{} - \alpha_3^{}\chi  \phi^T_{}i\tau_2^{}\xi_{2}^{} \phi
 - \mu \chi \xi_2^\dagger \xi_1\nonumber\\
&&  +\textrm{H.c.}\,.
\end{eqnarray}
After the singlet scalar $\sigma$ develops its $vev$ the first line
will induce the Lagrangian (\ref{eq:lagrangian2}), so that the
right-handed neutrinos obtain their Majorana masses. The second line
will generate the type-II seesaw in the presence of $\vev{\chi}$.

In our model, the global $U(1)_{lep}^{}$ is assumed to break at a very
large scale by $\vev{\chi} \sim 10^{13}_{}\,\textrm{GeV}$. The triplet
scalars $\xi_{1}^{}$ and $\xi_{2}^{}$, whose masses $\sim M_\xi^{}$
are of the order of $\vev{\chi}$, mix with each other and pick up tiny
$vev$s after the electroweak symmetry breaking,
\begin{eqnarray}
\vev{ \xi_2^{}} \sim -\alpha_{3}^{} {\langle \chi
\rangle\langle \phi \rangle^2_{} \over M_\xi^2}\,,~~~\langle
\xi_1^{} \rangle \sim -{\mu \langle \chi \rangle \langle \xi_2^{}
\rangle\over M_\xi^2} \,.
\end{eqnarray}
These triplet $vev$s will give rise to the left-handed neutrinos
Majorana mass matrix,
\begin{equation}
 m_{\nu_{ij}}^{} = f_{ij} \vev{ \xi_1^{}} \,.
\end{equation}

The CP-violating and out-of-equilibrium decays of the triplet scalars
$\xi_{1}$ and $\xi_{2}$ into the SM lepton and Higgs doublets can
generate a lepton asymmetry \cite{PhysRevLett.80.5716}. The sphaleron
\cite{kuzmin:1985mm} processes active in the range
$100\,\textrm{GeV}\lesssim T\lesssim 10^{12}_{}\,\textrm{GeV}$, will
partially convert this lepton asymmetry to a baryon asymmetry for
explaining the matter-antimatter asymmetry of the universe. %
In order to successfully induce leptogenesis and suppress washout
processes, we require that the singlet scalar $\sigma$ develops its
$vev$ after the sphaleron epoch is over, for example, we take
$\vev{\sigma}\sim 100\,\textrm{GeV}$. Through their Yukawa couplings
to $\sigma$ the right-handed neutrinos acquire their Majorana masses
$M = \alpha_1^{} \vev{ \sigma}$ and $\tilde M = \alpha_2^{} \vev{
  \sigma}$, expected to lie below $100\,\textrm{GeV}$ or so.

The spontaneous breakdown of the global lepton number global symmetry
through $\langle\chi\rangle$, leads to a Goldstone boson, whose
profile can be determined by the symmetry~\cite{schechter:1982cv},
leading to,
\begin{equation}
    {\cal G} = {1 \over N} \left[ \vev{ \chi} {\rm Im}
    (\chi)
    +
    \vev{ \xi_1^{}} {\rm Im} (\xi^0_1) +\vev{ \xi_2^{}} {\rm Im} (\xi^0_2)\right]
    \,.
\end{equation}
where $N$ is a suitable normalization, which is of the order of the
lepton number breaking scale $\sim\vev{\chi}$.  Clearly, the triplet
component of the Goldstone boson is highly suppressed by the ratio of
the triplet $vev$s over the singlet $vev$, suppressing its coupling to
the Z-bosons~\cite{schechter:1982cv}.

Note that since lepton number conservation forbids the couplings of
$\chi$ to the right-handed neutrinos, $\vev{\chi}$ will not have any
effect on them. The states $N_{R_1}^{}$ and $N_{R_2}^{}$ mix maximally
and in the basis where their Majorana mass matrix is diagonal, the
degenerate states $N_{R_{1}}^{}$ and $N_{R_{2}}^{}$ in
Eq. (\ref{eq:N12}) with mass $\pm M$ both couple to the left-handed
neutrinos through equal Yukawa couplings.  In addition, the state
$N_{R_3}^{}$ has no couplings to the left-handed neutrinos. Therefore,
the resulting neutrino mass matrix is a null matrix. This implies that
the strongest constraints on the couplings, for example those set by
the smallness of neutrino masses, are absent, very much like the case
of the inverse seesaw model~\cite{mohapatra:1986bd,Deppisch:2005zm}.

Another most severe constraint on the couplings of the right-handed
neutrinos with the charged leptons comes from their contribution to
$\mu \to e \gamma$, which is trivially removed now. Since here only
the $N_{R_1}^{}$ and $N_{R_2}^{}$ would mediate the process and they
are degenerate, the $\mu \to e \gamma$ amplitude would depend on the
effective light neutrino masses, which now vanishes, hence avoiding
the constraint on the couplings.

Note also that the physical Higgs boson can significantly decay into
$h \to \nu_{L}^{} + N_{R_1}^{}$ $h \to \nu_{L}^{} + N_{R_2}^{}$,
leading to a mono-jet-like signal.  In contrast, the decay $h \to
\nu_{L}^{} + N_{R_3}^{}$ does not take place.

The third right-handed neutrino $N_{R_3}^{}$ can not decay at all
because its only Yukawa coupling is with the singlet scalar
$\sigma$. This means that $N_{R_3}^{}$ will contribute a sizeable
relic density to the Universe. One can indeed check that $N_{R_3}^{}$
with the mass of a few GeV can have a desired cross section to serve
as the dark matter for $\vev{\sigma} \sim 100\,\textrm{GeV}$.

As a last comment we note that there may be a sizeable quartic
interaction between the singlet scalar $\sigma$ and the SM Higgs
doublet $\phi$, i.e.
$\lambda(\sigma^{\dagger}_{}\sigma)(\phi^\dagger_{}\phi)$. This term
can not be forbidden by imposing extra symmetries, and there is no
\emph{a priori} reason for $\lambda$ to be small.
In the presence of such coupling this dark matter $N_{R_3}^{}$ can be
searched in the decays of the Higgs boson produced at the LHC, $h \to
N_{R_3}^{} + N_{R_3}^{}$, resulting in a missing momentum signal.

\section{Summary}

Due to some special symmetry it may happen that the type-I seesaw
mechanism does not generate the observed neutrino masses, despite the
co-existence of sizeable Dirac mass terms and relatively low
right-handed neutrino Majorana masses. Such \emph{null seesaw
  mechanism} can be understood as a cancelation between the
contributions from the right-handed neutrinos. In this case there is
only a very mild constraint on the right-handed neutrinos, which can
have sizeable Yukawa couplings to the SM particles even if they are
light. The leading Higgs boson decay mode into a left-handed neutrino
and a right-handed neutrino could be probed at the LHC.  In the model
we have presented, one of the right-handed neutrinos has no Yukawa
couplings to the SM states, a fact that follows from our assumed
symmetry.  Hence it can provide the relic density required to solve
the puzzle of the dark matter. Finally, in our model the source of
neutrino masses is the type-II seesaw contribution arising from the
induced $vev$s of scalar Higgs triplets. The CP-violating and
out-of-equilibrium decays of these scalar triplets may also account
for the matter-antimatter asymmetry of the Universe through the
leptogenesis mechanism.

This work was supported by Spanish grants FPA2008-00319/FPA and
FPA2008-01935-E/FPA and ILIAS/N6 Contract RII3-CT-2004-506222.

 \def\baselinestretch{1}
\bibliographystyle{h-physrev4}

\begin{thebibliography}{10}


\bibitem[Neutrino (2008)]{Neutrino2008}
  See talks given at the
  Neutrino2008 conference Web-page\\
  http://www2.phys.canterbury.ac.nz/~jaa53/


\bibitem[Schwetz, T. et~al (2008)]{Schwetz:2008er}
  Schwetz, T., Tortola, M.~A., and Valle, J. W.~F.,
  arXiv:0808.2016 [hep-ph];
  for analysis details and the relevant experimental references see
  review by Maltoni, M., et~al., \emph{New J.  Phys.}, \textbf{6}, 122
  (2004).

\bibitem{Valle:2006vb}
For a review see J.~W.~F. Valle,
\newblock J. Phys. Conf. Ser. {\bf 53}, 473 (2006), [hep-ph/0608101],
\newblock  lectures at  Corfu Summer Institute on Elementary
  Particle Physics, Sept. 2005.

\bibitem{schechter:1980gr}
J.~Schechter and J.~W.~F. Valle,
\newblock Phys. Rev. {\bf D22}, 2227 (1980).

\bibitem{Minkowski:1977sc}
P.~Minkowski,
\newblock Phys. Lett. {\bf B67}, 421 (1977).
T.~Yanagida,
\newblock (KEK lectures, 1979),
\newblock ed. Sawada and Sugamoto (KEK, 1979).
M.~Gell-Mann, P.~Ramond and R.~Slansky,
\newblock (1979),
\newblock Print-80-0576 (CERN).
S.~Glashow,
\newblock (1980),
\newblock ed. M. Levy et al. (Plenum, New York), p. 707.
R.~N. Mohapatra and G.~Senjanovic,
\newblock Phys. Rev. Lett. {\bf 44}, 91 (1980).

\bibitem{schechter:1982cv}
J.~Schechter and J.~W.~F. Valle,
\newblock Phys. Rev. {\bf D25}, 774 (1982).

\bibitem{mohapatra:1986bd}
R.~N. Mohapatra and J.~W.~F. Valle,
\newblock Phys. Rev. {\bf D34}, 1642 (1986).

\bibitem{gonzalez-garcia:1989rw}
M.~C. Gonzalez-Garcia and J.~W.~F. Valle,
\newblock Phys. Lett. {\bf B216}, 360 (1989).

\bibitem{Deppisch:2005zm}
F.~Deppisch, T.~S. Kosmas and J.~W.~F. Valle,
\newblock Nucl. Phys. {\bf B752}, 80 (2006), [hep-ph/0512360];
F.~Deppisch and J.~W.~F. Valle,
\newblock Phys. Rev. {\bf D72}, 036001 (2005), [hep-ph/0406040], and
references therein.

\bibitem{Dittmar:1990yg}
M.~Dittmar {\em et~al.},
\newblock Nucl. Phys. {\bf B332}, 1 (1990).
M.~C. Gonzalez-Garcia, A.~Santamaria and J.~W.~F. Valle,
\newblock Nucl. Phys. {\bf B342}, 108 (1990).

\bibitem{delAguila:2008hw}
F.~del Aguila and J.~A. Aguilar-Saavedra,
\newblock 0809.2096.

\bibitem{Kersten:2007vk}
J.~Kersten and A.~Y. Smirnov,
\newblock Phys. Rev. {\bf D76}, 073005 (2007), [0705.3221].

\bibitem{deGouvea:2007uz}
A.~de~Gouvea,
\newblock 0706.1732.

\bibitem{Kim:2008pp}
Y.~G. Kim, K.~Y. Lee and S.~Shin,
\newblock JHEP {\bf 05}, 100 (2008), [0803.2932].

\bibitem{Gu:2007gy}
P.-H. Gu,
\newblock Phys. Lett. {\bf B661}, 290 (2008), [0710.1044].
\newblock Phys. Lett. {\bf B657}, 103 (2007), [0706.1946].

\bibitem{fukugita:1986hr}
M.~Fukugita and T.~Yanagida,
\newblock Phys. Lett. {\bf B174}, 45 (1986).

\bibitem{kuzmin:1985mm}
V.~A. Kuzmin, V.~A. Rubakov and M.~E. Shaposhnikov,
\newblock Phys. Lett. {\bf B155}, 36 (1985).

\bibitem{PhysRevLett.80.5716}
E.~Ma and U.~Sarkar,
\newblock Phys. Rev. Lett. {\bf 80}, 5716 (1998).



\end{thebibliography}

\end{document}